\documentclass[12pt,a4paper]{article}

\newcommand{\beq}{\begin{equation}}

\newcommand{\eeq}{\end{equation}}
\newcommand{\bea}{\begin{eqnarray}}
\newcommand{\eea}{\end{eqnarray}}
\newcommand{\rf}[1]{(\ref{#1})}

\newcommand{\Tr}{\mbox{Tr}}

\begin{document}
\input{epsf}
\topmargin 0pt
\oddsidemargin 5mm
\headheight 0pt
\headsep 0pt
\topskip 9mm
\pagestyle{empty}

\begin{center}

\begin{flushright}
{\sc\footnotesize hep-th/0402033}\\
{\sc\footnotesize NORDITA-2004-3}\\
\end{flushright}
\vspace*{100pt}

{\large \bf {Three-spin Strings 
on AdS$_5$ $\times$ S$^5$ from 
${\cal N}=4$ SYM}}
\newline
\vspace{26pt}

{\sl C.\ Kristjansen}
\\
\vspace{6pt}
{\sl NORDITA,} \\
{\sl Blegdamsvej 17, DK-2100 Copenhagen \O}\\

\vspace{20pt}
\end{center}

\begin{abstract}

Using the integrable spin chain picture we study the one-loop anomalous
dimension of certain single trace scalar operators of ${\cal N}=4$ SYM
expected to correspond to semi-classical string states on $AdS_5\times S^5$
with three large angular momenta
$(J_1,J_2,J_3)$ on $S^5$. In particular, we investigate the analyticity 
structure encoded in the Bethe equations for various distributions of 
Bethe roots.
 In a certain region of the parameter space
our operators reduce to the gauge theory duals of 
the {\it folded} string with two large 
angular momenta and in another region to the duals of
the {\it circular} string with angular momentum assignment $(J,J',J')$, $J>J'$.
In between we locate a critical line. We propose that the operators
above the critical line are the gauge theory duals of the circular elliptic
string with three different spins and support this by a perturbative
calculation.
\end{abstract}

\newpage

\pagestyle{plain}

\setcounter{page}{1}

\section{Introduction}
Recent development, triggered by the pp-wave/BMN 
correspondence~\cite{Berenstein:2002jq}, has led to new insights on the 
duality between string theory and gauge theory and has in particular 
revealed interesting novel integrability structures in both kind of
theories~\cite{Minahan:2002ve,Beisert:2003tq,Beisert:2003jj,Beisert:2003yb,
Mandal:2002fs,Bena:2003wd,Dolan:2003uh,Alday:2003zb,Arutyunov:2003uj,
Arutyunov:2003za}. The progress 
has been made by focusing on a simple set of observables which, 
according
to the original AdS/CFT
correspondence~\cite{Maldacena:1997re}, should be closely
related, namely the energy spectrum of 
single string states on AdS$_5\times S^5$ and anomalous dimensions of 
local single trace operators of ${\cal N}=4$ SYM. Here the string states
are characterized by various quantum numbers such as
angular momenta
and these should match the representation labels of the 
corresponding operators. Following the formulation of the pp-wave/BMN
correspondence efficient techniques for evaluating anomalous 
dimensions of ${\cal N}=4$ SYM operators were 
developed~\cite{Kristjansen:2002bb}.
A further crucial step was the discovery of 
Minahan and Zarembo that the planar one-loop dilatation operator in
the scalar sector of ${\cal N}=4$ SYM could be identified as the
Hamiltonian of an integrable $SO(6)$ spin chain~\cite{Minahan:2002ve}.
This conquest was later extended to the set of all operators
in ${\cal N}=4$ SYM and yielded an integrable $PSU(2,2|4)$ super spin 
chain~\cite{Beisert:2003jj,Beisert:2003yb}. 
A similar spin chain picture, relating to space-time- rather than
internal symmetries, was previously discovered in the context of
QCD~\cite{Lipatov:1993yb}.
On the string theory side
it was realized that the classical energy of certain string states
with several large angular momenta on $S^5$
exhibited an analytical dependence on the parameter
$\lambda'=\frac{\lambda}{J^2}$ where $J$ is the total 
angular momentum  and where
$\lambda$ is the squared string 
tension which, via the AdS/CFT dictionary, is mapped onto the 't Hooft
coupling constant $g_{\mbox{\tiny YM}}^2 N$.
Furthermore, for these states
string quantum corrections were suppressed as $\frac{1}{J}$ in the
limit $J\rightarrow \infty$, $\lambda'=\frac{\lambda}{J^2}$ 
fixed~\cite{Frolov:2003qc,Frolov:2003tu}\footnote{Several other types
of string states with similar properties have been found. These include
string states with non-vanishing angular momenta on
$AdS_5$~\cite{Frolov:2003qc,Arutyunov:2003uj,Arutyunov:2003za} as well
as a class of  so-called pulsating string
solutions~\cite{Minahan:2002rc,Engquist:2003rn}.}.
This led to the suggestion
that taking the limit $J\rightarrow \infty$ for fixed $\lambda'$,
the term linear in $\lambda'$ in the small $\lambda'$
expansion of the classical string energy 
would match the one-loop anomalous dimension of a gauge theory
operator carrying the same (large) $SO(6)$
representation labels as the string
state~\cite{Frolov:2003qc,Frolov:2003tu}. In the spin chain picture,
determining the one-loop anomalous dimension amounts to solving a set
of Bethe equations~\cite{Bethe:1931hc} and considering large
representation labels (i.e.\ long operators) corresponds to going
to the thermodynamical limit.

Comparison of classical string energies and one-loop anomalous
dimensions  has been successfully carried out in a number of cases, the
prime example involving strings  carrying two large angular momenta
$(J_1,J_2)$ on $S^5$. On the gauge theory side two types of solutions
of the Bethe equations were found~\cite{Beisert:2003xu} and these
were identified as the gauge theory dual of respectively a folded and
a circular string~\cite{Beisert:2003xu,Frolov:2003xy} with
the folded string being the
one of lowest energy. Expressions giving the one-loop anomalous
dimension, respectively the ${\cal O}(\lambda')$ contribution to the
classical energy as a function of the representation labels in
a parametric form were found. These parametrizations involved
elliptic integrals and were shown to match at a functional 
level~\cite{Beisert:2003ea,Arutyunov:2003uj}. The situation where
three spins $(J_1,J_2,J_3)$ on $S^5$ are non-vanishing is less well
understood. One particularly simple three spin circular string solution was
found a while ago~\cite{Frolov:2003qc,Frolov:2003tu}. It has two of
its three spins equal, i.e.\ $J_2=J_3$ and is stable for large enough
values of $J_1$. This solution is again parametrized in terms of
elliptic functions and its gauge theory dual was identified 
in~\cite{Engquist:2003rn}. The generic case of (rigid) strings with
three {\it different} $S^5$ spin quantum numbers was studied 
in~\cite{Arutyunov:2003uj} where it was shown that the relevant
sub-sector of the string $\sigma$-model could be mapped onto an
integrable Neumann model. Further generalizations and relations to 
integrable models were found in~\cite{Arutyunov:2003za}. The parallel
gauge theory analysis is so far lacking. A characteristic feature
which distinguishes the three-spin solutions from the two-spin
ones is that whereas the latter are conveniently parametrized in terms
of elliptic integrals the former generically
require the use of {\it hyper-}elliptic 
integrals. There does, however, exist a class of three-spin solutions
which are still elliptic~\cite{Arutyunov:2003uj}. 
In reference~\cite{Arutyunov:2003uj} particular attention
was paid to hyper-elliptic 
three spin solutions generalizing respectively the
folded and circular two-spin string. Of these three-spin solutions 
the circular one exists in an elliptic version
whereas the folded one does not~\cite{Arutyunov:2003uj}.
Here, we shall study a class of
holomorphic gauge theory operators carrying generic $SO(6)$
representation labels $(J_1,J_2,J_3)$.
In a certain region of the parameter space (corresponding to $J_3=0$)
the operators reduce to the gauge theory duals of the {\it folded}
two-spin string whereas in another one they constitute the duals of the
{\it circular} string with spin assignment $(J,J',J')$,
$J>J'$~\cite{Engquist:2003rn}.
We will show that these two
different manifestations of the dual string are made possible through
the existence of a line of critical points in the parameter space. 
Furthermore, we propose that above the critical line 
the gauge theory operators studied are the duals of the
circular elliptic three-spin states of~\cite{Arutyunov:2003uj}. 
The proposal is supported by a 
perturbative calculation.

\section{The general gauge theory set-up \label{setup}}

Gauge theory operators dual to rigid strings 
with three non-vanishing angular momenta, $(J_1,J_2,J_3)$, on $S^5$
are expected to be operators of the type \\
$\Tr( (\chi \chi)^k X^{J_1-k} Y^{J_2-k} Z^{J_3-k} +perm's)$, 
$ k< \min\{J_1,J_2,J_3\}$,
where $X$, $Y$ and $Z$ are the three complex scalars of ${\cal N}=4$ 
SYM with $SO(6)$
weights $(1,0,0)$, $(0,1,0)$ and $(0,0,1)$ and where $\chi$ is the fermion with
$SO(6)$ weight $(1/2,1/2,1/2)$. In the present paper we shall work at
one-loop order, i.e. at ${\cal O}(\lambda)$, where the 
dilatation generator only
mixes the operators without fermionic constituents. We shall thus be interested
in diagonalizing the one-loop dilatation generator in the sub-set of
operators of the type $\Tr(X^{J_1} Y^{J_2} Z^{J_3} + perm's)$ or equivalently
diagonalizing the Hamiltonian of the integrable $SO(6)$ spin chain in the
appropriate sub-set of spin states.
%Gauge theory operators dual to rigid strings 
%with three non-vanishing angular momenta, $(J_1,J_2,J_3)$, on $S^5$
%are expected to be operators of the
%type $\Tr(X^{J_1} Y^{J_2} Z^{J_3}+ perm's)$ where $X$, $Y$ and $Z$ are the 
%three complex scalars of ${\cal N}=4$ SYM. We are therefore
%interested in diagonalizing
%the one-loop dilatation generator of ${\cal N}=4$ SYM in the subset of such
%operators --- or equivalently diagonalizing the Hamiltonian of the 
%integrable $SO(6)$ spin chain in the appropriate subset of spin states. 
 The spin chain picture is particularly
convenient when considering operators for which $L\equiv J_1+J_2+J_3\rightarrow
\infty$. Finding an eigenstate and corresponding eigenvalue of the 
$SO(6)$ spin chain Hamiltonian consists in solving a set of 
algebraic equations for
the Bethe roots. For the $SO(6)$ spin chain there are three
different types of Bethe roots reflecting the fact that
the Lie algebra $SO(6)$ has three simple roots. 
However, for holomorphic operators only two of the three types of roots can
be excited.
Denoting the number of 
roots of the two relevant types as $n_1$ and $n_2$ and the
roots themselves as $\{u_{1,j}\}_{j=1}^{n_1}$ and $\{u_{2,j}\}_{j=1}^{n_2}$ 
the Bethe equations read
\bea
\left(\frac{u_{1,j}+i/2}{u_{1,j}-i/2}\right)^L
&=&\prod_{k\neq j}^{n_1} 
\frac{u_{1,j}-u_{1,k}+i}{u_{1,j}-u_{1,k}-i}
\prod_{k=1}^{n_2} \frac{u_{1,j}-u_{2,k}-i/2}{u_{1,j}-u_{2,k}+i/2},
\label{Bethe1} \\
1&=&\prod_{k\neq j}^{n_2} 
\frac{u_{2,j}-u_{2,k}+i}{u_{2,j}-u_{2,k}-i}
\prod_{k=1}^{n_1} \frac{u_{2,j}-u_{1,k}-i/2}{u_{2,j}-u_{1,k}+i/2}.
\label{Bethe2}
\eea
We shall assume that $n_1\leq \frac{L}{2}$, $n_2\leq \frac{n_1}{2}$. 
The $SO(6)$ representation implied by this choice of Bethe roots is given
by the Dynkin labels $[n_1-2n_2,L-2n_1+n_2,n_1]$.  In terms of
the spin quantum numbers, assuming $J_1\geq J_2\geq J_3$ this
corresponds to $[J_2-J_3,J_1-J_2,J_2+J_3]$ or $J_1=L-n_1$, $J_2=n_1-n_2$,
$J_3=n_2    $. A given solution of the Bethe equations gives rise to an
eigenvalue of the spin chain Hamiltonian i.e.\ a one loop anomalous
dimension which is
\beq
\label{gammafirst}
\gamma=\frac{\lambda}{8\pi^2}\sum_{j=1}^{n_1}\frac{1}{(u_{1,j})^2+1/4}.
\eeq
To enforce the cyclicity of the trace we have in addition 
to the equations~\rf{Bethe1} and~\rf{Bethe2} the following constraint
\beq
1=\prod_{j=1}^{n_1}\left(\frac{u_{1,j}+i/2}{u_{1,j}-i/2}\right).
\label{cyclicity}
\eeq
In the thermodynamical limit $L\rightarrow\infty$ all roots are ${\cal O}(L)$
and it is convenient to re-scale them accordingly. Doing so, taking
the logarithm of the Bethe equations and imposing the limit $L\rightarrow
\infty$ one is left with a set of integral equations.

\section{The present gauge theory set-up}
Let us define
\beq
\alpha=\frac{n_1}{L},
\hspace{0.7cm} \beta=\frac{n_2}{L}.
\eeq
Then the spin quantum numbers are given by
$(J_1,J_2,J_3)=((1-\alpha)L,(\alpha-\beta)L,\beta L)$.
We shall assume that the Bethe
roots $\{u_{1,j}\}_{j=1}^{n_1}$ are distributed as in the case of the
folded two spin string 
solution of reference~\cite{Beisert:2003xu}, i.e.\ they live
on two arches in the complex plane, ${\cal C}_+$ and ${\cal C}_-$, which are
each others mirror images with respect to zero. Each arch is symmetric around
the real axis and neither one intersects the imaginary axis. 
For this configuration the constraint~\rf{cyclicity}
is fulfilled (but $n_1$ is required to be even).
Furthermore, let us assume that the roots $\{u_{2,j}\}_{j=1}^{n_2}$
live on some curve ${\cal C}_2$ not intersecting ${\cal C}_+$ or ${\cal C}_-$.

Performing the above mentioned manipulations relevant for the thermodynamical
limit we can write the two Bethe equations 
as in~\cite{Engquist:2003rn}
\bea
\frac{1}{u} - 2\pi m &=& 2
\int_{{\cal C}_+}\hspace*{-0.6cm}-\hspace*{0.2cm}
du'\frac{\sigma(u')}{u-u'}
+ 2 \int_{{\cal C}_+}du'\frac{\sigma(u')}{u+u'}
- \int_{{\cal C}_2}du'\frac{\rho_2(u')}{u-u'},\hspace{0.3cm}
u\in {\cal C}_+,
\label{rootone} \\
2 \pi m_2& =& 2 \int_{{\cal C}_2}\hspace*{-0.55cm}-\hspace*{0.25cm}
du' \frac{\rho_2(u')}{u-u'}-
 \int_{{\cal C}_+}du'\frac{\sigma(u')}{u-u'}
-  \int_{{\cal C}_+}du'\frac{\sigma(u')}{u+u'},
\hspace{0.3cm}
u\in {\cal C}_2,
\label{root2}
\eea
where $m$ and $m_2$ are integers which reflect the ambiguities coming 
from the different possible choices of branches for the logarithm and
where $\int\hspace*{-0.35cm}-$ means that the integral
has to be understood in the principal value sense.
Furthermore, $\rho_2(u)$ and $\sigma(u)$ are root densities 
describing the continuum distribution of $\{u_{2,j}\}_{j=1}^{n_2}$
and the subset of $\{u_{1,j}\}_{j=1}^{n_1}$ with positive real part,
respectively. The densities are 
normalized as
\beq
\frac{\alpha}{2}=\int_{{\cal C}_+} \sigma(u)du, \hspace{1.0cm}
\beta=\int_{{\cal C}_2} \rho_2(u)du.
\eeq
We shall shortly see that the mode number $m_2$ actually has to 
vanish\footnote{This is natural from the spin chain point of view
as $m_2$ can be interpreted as a discrete momentum associated with
the roots $\{u_{2,j}\}_{j=1}^{n_2}$ 
and all momentum is known to be carried by
the roots $\{u_{1,j}\}_{j=1}^{n_1}$ (cf.\ eqn.~\rf{gammafirst}).}.
Rather than working with the densities we prefer to work with the
resolvents $W(u)$ and $W_2(u)$ defined by
\beq \label{resolvents}
W(u)=\int_{{\cal C}_+} du'\frac{\sigma(u')}{u-u'}, \hspace{1.0cm}
W_2(u)=\int_{{\cal C}_2} du'\frac{\rho_2(u')}{u-u'}.
\eeq
The resolvents are analytic in the complex plane except for a cut
respectively along ${\cal C}_+$ and ${\cal C}_2$. 
In the continuum language the one-loop anomalous dimension, $\gamma$, is 
 given by
\beq
\gamma=\frac{\lambda}{4\pi^2 L}\int_{{\cal C}_+} du\, \frac{\sigma(u)}{u^2}=
-\frac{\lambda}{4\pi^2 L} W'(0). \label{gammacont}
\eeq
Not only are the resolvents technically more convenient. It appears that they
are indeed objects with a direct physical interpretation. For instance, $W(u)$
is the generating function of all the higher conserved charges of the spin
chain~\cite{Arutyunov:2003rg}. It would be interesting to gain a similar
understanding of $W_2(u)$.

One possible configuration for the roots $\{u_{2,j}\}_{j=1}^{n_2}$
 is that they lie in an
interval $[-ic,ic]$ on the imaginary axis~\cite{Engquist:2003rn}. 
In reference~\cite{Engquist:2003rn} the case $c\rightarrow \infty$
was studied and the corresponding string state was identified as the
circular string of~\cite{Frolov:2003qc} with spin assignment
$(J,J',J')$,
$J>J'$.
Here
we shall analyze the generic $c$ case.
Our strategy when solving
the Bethe equations will be the same as that of 
reference~\cite{Engquist:2003rn}. We will express $\rho_2(u)$ in terms
of $\sigma(u)$ by means of eqn.~\rf{root2} and use the resulting expression
to eliminate $\rho_2(u)$ from eqn.~\rf{rootone}. We see that $\rho_2(u)$
only enters eqn.~\rf{rootone} 
via the corresponding resolvent.
Thus we do not
need to determine $\rho_2(u)$ itself. Rewriting eqn.~\rf{root2} as
\beq
\int_{{\cal C}_2}
\hspace*{-0.55cm}-\hspace*{0.25cm} du' \frac{\rho_2(u')}{u-u'}
=\pi m_2
+\int_{{\cal C}_+}du'\frac{\sigma(u') u}{u^2-u'^2}, \hspace{0.4cm}
u\in {\cal C}_2,
\eeq
we recognize the saddle point equation of the Hermitian one-matrix model with
the terms on the right hand side playing the role of the derivative of
the potential. Thus we
can immediately write down a contour integral expression for the resolvent, 
see f.\ inst.~\cite{Ambjorn:1992gw}
\beq
W_2(u)
= \oint_{\cal C} \frac{d\omega}{2 \pi i} \frac{1}{u-\omega} 
\sqrt{\frac{u^2+c^2}{\omega^2+c^2}}\left\{ \pi m_2 +
 \int_{{\cal C}_+} du'\frac{\sigma(u') \omega}{\omega^2-
u'^2} \right\},
\eeq
where the contour ${\cal C}$ encircles the interval $[-ic,ic]$ but not the 
various other singularities
of the integrand. Interchanging the order of integrations in the 
last term we can write this as
\beq
W_2(u)=m_2 \pi +
 \int_{{\cal C}_+}
\hspace*{-0.65cm}-\hspace*{0.15cm} du'\frac{\sigma(u')u}{u^2-u'^2}
- \int_{{\cal C}_+} \hspace*{-0.65cm}-\hspace*{0.15cm} 
du' \frac{u'\sigma(u')}{u^2-u'^2}
\sqrt{\frac{u^2+c^2}{u'^2+c^2}}.
\label{W2}
\eeq
The parameter $c$ can be expressed in terms of $\alpha$ and $\beta$ by
making use of the
asymptotic
behaviour of $W_2(u)$ as $u\rightarrow \infty$. One has
\beq
W_2(u)\sim \frac{\beta}{u},\hspace{0.5cm} \mbox{as} 
\hspace{0.5cm}u\rightarrow \infty,
\eeq
which immediately gives
\bea
0&=&m_2\pi, \label{bound1} \\
\beta&=& \frac{\alpha}{2}-  \label{bound2}
\int_{{\cal C}_+}du \frac{\sigma(u) u}{\sqrt{u^2+c^2}}.
\eea
We notice the following two limiting cases of eqn.~\rf{W2}
which serve as a consistency check
of our solution
\beq
\lim_{c\rightarrow 0} W_2(u) =0,
\eeq
\beq
\lim_{c\rightarrow \infty} W_2(u) = 
\int_{{\cal C}_+}\hspace*{-0.65cm}-\hspace*{0.15cm} 
du'\frac{\sigma(u')}{u+u'}.
\eeq
Here the last expression coincides with the one obtained in 
reference~\cite{Engquist:2003rn}. 
As noted in respectively~\cite{Beisert:2003xu} and~\cite{Engquist:2003rn}
the integral equation~\rf{rootone} reduces to that of the $O(n)$ model on
a random lattice~\cite{Kostov:fy} 
with $n=-2$ for $c\rightarrow 0$ and
$n=-1$ for $c\rightarrow \infty$. The $O(n)$ model on a random
lattice can be solved exactly
for any value of $n$ and the solution is for generic $n$ parametrized in
terms of elliptic functions~\cite{Eynard:1995nv}. 
However, a simplification
occurs at the so-called rational points where $n=2\cos(\pi \frac{p}{q})$
with $p$ and $q$ co-prime integers~\cite{Kostov:pn,Eynard:1992cn}.
The reason why elliptic integrals appear can most easily be understood by
rewriting the integral equation of the $O(n)$ model in terms of the 
resolvent $W(u)$
which, as mentioned above,
 is analytic in the complex plane except for a cut along the
contour ${\cal C}_+$. The relevant integral equation involves $W(u)$ as
well as $W(-u)$. Effectively, one thus has {\it two} cuts and that
is what leads to the elliptic structure for generic values of $n$.
For details we refer to~\cite{Eynard:1995nv}.

We can conveniently rewrite the expression~\rf{W2} for $W_2(u)$ as 
\[
W_2(u)=
\frac{1}{2}\int_{{\cal C}_+}
\hspace*{-0.65cm}-\hspace*{0.15cm} du' \frac{\sigma(u')}{u-u'}
\left(1-\sqrt{\frac{u^2+c^2}{u'^2+c^2}}\right)
+\frac{1}{2}\int_{{\cal C}_+}\hspace*{-0.65cm}-\hspace*{0.15cm}
du' \frac{\sigma(u')}{u+u'}
\left(1+\sqrt{\frac{u^2+c^2}{u'^2+c^2}}\right).
\]
Inserting this expression for $W_2(u)$ in eqn.~\rf{rootone} we get
the following integral equation for general $c$
\bea
\lefteqn{\frac{1}{u}-2\pi m= 
\label{finalrootone} }\\ 
&& \frac{1}{2}\int_{{\cal C}_+} \hspace*{-0.6cm}-\hspace*{0.3cm}
du'\frac{\sigma(u')}{u-u'}
\left(3+\sqrt{\frac{u^2+c^2}{u'^2+c^2}}\right)
+\frac{1}{2}\int_{{\cal C}_+}du' \frac{\sigma(u')}{u+u'}
\left(3-\sqrt{\frac{u^2+c^2}{u'^2+c^2}}\right),
\nonumber
\eea
with $u\in {\cal C}_+$.
We can trade the square roots in eqn.~\rf{finalrootone} 
for extra poles (or rather cuts) by
performing a change of variables,
obtaining an integral equation which 
exposes the analyticity structure of the problem in a simpler manner.
The relevant changes of variables are different for small and for
large $c$ and the resulting integral equations show that there is a
phase transition taking place at some intermediate value of $c$. This
explains why the string state dual to the operator
considered does not need to be of the same type for
$c\rightarrow 0$ (folded) as for $c\rightarrow \infty$ (circular).

\section{The case of small $c$}
For $c$ small a convenient change of variables is
\beq
u=\frac{p^2+c^2}{2ip}, \hspace{0.7cm} 
u'=\frac{q^2+c^2}{2iq}, \label{change2}
\eeq
which is well-defined as $c\rightarrow 0$ but not as $c\rightarrow \infty$.
With this change of variables we get
\beq
\sqrt{\frac{c^2+u^2}{c^2+u'^2}}=\frac{q(p^2-c^2)}{p(q^2-c^2)},
\eeq
and we see that  the limit
$c\rightarrow 0$ is as we wish. Inserting the change of variables~\rf{change2}
into the integral equation~\rf{finalrootone} we get with $du \sigma(u) \equiv
dq \rho(q)$
\beq
\frac{p}{p^2+c^2}+i\pi m= \label{intsmallc}
\int_{\tilde{\cal C}_+} 
\hspace*{-0.6cm}-\hspace*{0.3cm}
dq \rho(q) \frac{q^2}{q^2-c^2}
\left\{\frac{p}{c^2-qp}+\frac{2}{p-q}+\frac{2}{p+q}+\frac{p}{c^2+qp}\right\},
\eeq
with $p\in {\tilde{\cal C}}_+$ where ${\tilde{\cal C}}_+$ is the
contour for the transformed roots $q$. 
The boundary equation~\rf{bound2} turns into
\beq
\beta=\frac{\alpha}{2}-\int_{\tilde{{\cal C}}_+}
dq \rho(q) \frac{q^2+c^2}{q^2-c^2},
\label{boundsmallc}
\eeq
and the expression for $\gamma$ becomes
\beq
\gamma=-\frac{\lambda }{\pi^2 L}
\int_{\tilde{{\cal C}}_+} dq \frac{\rho(q) q^2}{(q^2+c^2)^2}.
\eeq
Here it is convenient to define a resolvent by
\beq
W(p)=\int_{\tilde{{\cal C}}_+}dq \rho(q)\frac{q^2}{q^2-c^2}\frac{1}{p-q}.
\eeq
Again, $W(p)$ is analytic
in the complex plane except for a cut along the contour $\tilde{{\cal C}}_+$
and we can express the anomalous dimension, $\gamma$ through
$W(p)$ as
\beq
\gamma=-
\left.\frac{\partial}{\partial p^2} \left(\frac{p^2-c^2}{2p}
(W(p)-W(-p))\right)\right|_{p=ic}.
\eeq
 Apart from the function $W(p)$ the integral equation~\rf{intsmallc}
involves $W(-p)$, $W(\frac{c^2}{p})$ and $W(-\frac{c^2}{p})$. This integral
equation can be viewed as 
a ``super-position'' of that of the usual $O(n)$ model on a random 
lattice~\cite{Kostov:fy} and that of the plaquette model studied 
in~\cite{Chekhov:1996xy}. 
In particular, we see that we effectively have four different cuts. In other
words, the presence of the Bethe roots $\{u_{2,j}\}_{j=1}^{n_2}$
has the effect of 
introducing an extra pair of ``mirror'' cuts in the integral equation
for the Bethe roots $\{u_{1,j}\}_{j=1}^{n_1}$. 
Denoting the end points of the cut 
${\tilde{\cal C}}_+$ as $a$ and $b=-a^*$ and writing symbolically
${\tilde{\cal C}}_+=[a,b]$ (knowing that ${\tilde{\cal C}}_+$
is not a straight line) the other cuts are $[-b,-a]$, 
$[\frac{c^2}{a},\frac{c^2}{b}]$ and $[-\frac{c^2}{b},-\frac{c^2}{a}]$.
Such a 4-cut integral equation  
generically has a solution in terms of
hyper-elliptic integrals. However, since the weight of the additional
cuts can be written in the form $n=2\cos(p\pi/q)$ with $p$ and $q$
co-prime integers ($p=1$, $q=3$) we expect to have a situation which
generalizes the above mentioned rational points of the $O(n)$ model
on a random lattice. This indicates that the solution can be 
at most elliptic. 
As the present parametrization is designed to study the system for small 
values of $c$ we can assume that $|c|<|a|=|b|$. Then the cuts 
$[\frac{c^2}{a},\frac{c^2}{b}]$ and $[-\frac{c^2}{b},-\frac{c^2}{a}]$
are ``inside'' (i.e.\ closer to the origin than) the cuts $[a,b]$ and
$[-b,-a]$. When $c\rightarrow 0$ these inner cuts shrink to zero and 
disappear. In this limit we recover the $O(n=-2)$ model of 
reference~\cite{Beisert:2003xu}. 
As  $|c|\rightarrow |a|$ (or $\beta \rightarrow (\beta_c(\alpha))_-)$ the
two sets of cuts approach each other and a singularity occurs. 
The equation~\rf{intsmallc} looses its meaning, an obvious sign being
the divergence of the pre-factor $q^2/(q^2-c^2)$. As mentioned above,
this explains why the string state dual to the gauge theory
operator considered does not need to be of the same type for small and for 
large $c$.

\section{The case of large $c$}

To study the case where $c$ is large, let us return to eqn.~\rf{finalrootone}
and choose another change of variables. In this case we set
\beq
u=\frac{2ip}{1+\frac{p^2}{c^2}},\hspace{0.7cm} 
u'=\frac{2iq}{1+\frac{q^2}{c^2}},\label{change1}
\eeq
which we notice is well-behaved as $c\rightarrow\infty$ but singular as
$c\rightarrow 0$. Now, we find
\beq
\sqrt{\frac{c^2+u^2}{c^2+u'^2}}
=\frac{(1-\frac{p^2}{c^2})(1+\frac{q^2}{c^2})}
{(1+\frac{p^2}{c^2})(1-\frac{q^2}{c^2})}.
\label{squareroot}
\eeq
In accordance with the remark just above, this formula gives rise to the
correct asymptotic expansion as $c\rightarrow\infty$ but not as 
$c\rightarrow 0$. In the new variables the integral equation~\rf{finalrootone}
reads, with $du\,\sigma(u)\equiv dq\,\rho(q)$
\bea
\lefteqn{
\frac{1+\frac{p^2}{c^2}}{2p}+2\pi m i=} 
\label{intlargec}\\
&&\frac{1}{2}\int_{{\tilde{\cal C}}_+} 
\hspace*{-0.6cm}-\hspace*{0.3cm}
dq \rho(q)
\left(\frac{1+\frac{q^2}{c^2}}{1-\frac{q^2}{c^2}}\right)
\left\{\frac{1-\frac{qp}{c^2}}{p+q}+
\frac{\frac{2}{c^2}(p-q)}{1+\frac{qp}{c^2}}
+\frac{\frac{1}{c^2}(p+q)}{1-\frac{qp}{c^2}}
+\frac{2\left(1+\frac{qp}{c^2}\right)}{p-q}\right\},
\nonumber
\eea
where $p\in {\tilde{\cal C}}_+$ with ${\tilde{\cal C}}_+$ being the contour
for the transformed roots $q$. 
The boundary equation~\rf{bound2} turns into
\beq
\beta=\frac{\alpha}{2} -\frac{1}{c}
\int_{\tilde{{\cal C}}_+} dq \rho(q) \frac{2 i q}{1-\frac{q^2}{c^2}},
\eeq
and the expression for $\gamma$ reads
\beq
\gamma=-\frac{\lambda}{16 \pi^2 L} \int_{\tilde{\cal C}_+}
dq\rho(q)\frac{(1+q^2/c^2)^2}{q^2}.
\eeq
This time a natural definition of the resolvent is
\beq
W(p)=\int_{\tilde{{\cal C}}_+}
dq \rho(q)\left(\frac{1+\frac{q^2}{c^2}}{1-\frac{q^2}{c^2}} \right)
\frac{1+\frac{qp}{c^2}}{p-q},
\eeq
and $\gamma$ can be expressed as
\beq
\gamma=\left.\frac{\lambda}{16 \pi^2 L} \,
\frac{\partial}{\partial p}\left(
W(p)-W(-\frac{c^2}{p})\right) \right|_{p=0}.
\eeq
Once again, apart from $W(p)$ the integral equation involves $W(-p)$,
$W(\frac{c^2}{p})$ and $W(-\frac{c^2}{p})$. Hence,
we discover that the effect of the Bethe roots
$\{u_{2,j}\}_{j=1}^{n_2}$ has been to introduce an extra pair of
``mirror cuts'' in the integral equation for $\{u_{1,j}\}_{j=1}^{n_1}$
so that the density $\rho(q)$ now effectively
obeys a 4-cut integral equation. Also in this case the integral
equation shares some features with both the one of the $O(n)$ model on
a random lattice~\cite{Kostov:fy} and the one of the plaquette model
of~\cite{Chekhov:1996xy}.
Furthermore, due to the weights of the various cuts 
we expect to have a situation which generalizes the rational points
of the $O(n)$ model and thus a solution which is at most elliptic.
Denoting the end points of the cut
${\tilde{\cal C}}_+$ as $a$ and $b=-a^*$ and writing symbolically
${\tilde{\cal C}}_+=[a,b]$ (still knowing that ${\tilde{\cal C}}_+$ is not a 
straight line) the other cuts are $[-b,-a]$, $[\frac{c^2}{a},\frac{c^2}{b}]$
and $[-\frac{c^2}{b},-\frac{c^2}{a}]$. 
The present parametrization
is designed to study the case where $c$ is large. Therefore, let us consider
$|c|>|a|=|b|$. In this case the cuts $[\frac{c^2}{a},\frac{c^2}{b}]$
and $[-\frac{c^2}{b},-\frac{c^2}{a}]$ are ``outside'' (i.e. further from
the origin than) the cuts $[a,b]$ and $[-b,-a]$. When $c\rightarrow\infty$
the two outer cuts move out to infinity and disappear. In this limit
we recover the simple $O(n=-1)$ integral equation studied in 
reference~\cite{Engquist:2003rn}. When $|c|\rightarrow |a|$ 
(or $\beta\rightarrow (\beta_c(\alpha))_+$) the two 
sets of cuts approach each other and for $|c|=|a|$ a singularity occurs.
This coincides with the
divergence of the pre-factor $(1+q^2/c^2)/(1-q^2/c^2)$. 

\section{Perturbative expansion for $\beta\approx \frac{\alpha}{2}$ }

Let us define
\beq
\epsilon=\frac{\alpha}{2}-\beta,
\eeq
and let us consider $\epsilon\ll \alpha,\beta$.
In terms of angular
momenta we have
$(J_1,J_2,J_3)=((1-\alpha)L,(\frac{\alpha}{2}+\epsilon)L,
(\frac{\alpha}{2}-\epsilon)L)$ or
\beq
\epsilon=\frac{1}{2L}(J_2-J_3), \hspace{0.7cm} J_1> J_2,J_3.
\eeq
The operator in question is expected to be the gauge theory dual of
a slightly perturbed version of the circular three-spin
state of~\cite{Frolov:2003qc,Frolov:2003tu} which has angular momenta 
$(J,J',J')$, $J>J'$. Obviously, a small value of $\epsilon$ 
corresponds to a large value of $c$. Expanding the expression~\rf{W2} for
large $c$ we get
\beq
W_2(u)=\int_{{\cal C}_+}\hspace*{-0.6cm}-\hspace*{0.2cm}
du'\frac{\sigma(u')}{u+u'}
-\frac{1}{2c^2}\int_{{\cal C}_+} du' u' \sigma(u').
\eeq
Inserting this into the integral equation~\rf{rootone} and making use
of the boundary equation~\rf{bound2} gives
\beq
\frac{1}{u}-2 \pi m - \frac{\epsilon}{2c}=
 2\int_{{\cal C}_+}\hspace*{-0.6cm}-\hspace*{0.2cm}
du'\frac{\sigma(u')}{u-u'}
+  \int_{{\cal C}_+}du'\frac{\sigma(u')}{u+u'},
\hspace{0.7cm}u\in {{\cal C}_+}.
\eeq
This equation can again be recognized as the saddle point equation of 
the $O(n)$ model on a random lattice for $n=-1$, with the terms
on the left hand side playing the role of the derivative of the potential.
In terms of the resolvent of eqn.~\rf{resolvents} the equation reads
\beq
W(u+i0)+W(u-i0)-W(-u)=V'(u), \hspace{0.7cm} u\in {\cal C}_+,
\eeq
where
\beq
V'(u)=\frac{1}{u}-2\pi\left( m +\frac{\epsilon}{4\pi c}\right).
\eeq
The asymptotic behaviour of $W(u)$ is
\beq
W(u)\sim \frac{\alpha}{2u}+\frac{\epsilon \,c}{u^2}, \hspace{0.7cm} \mbox{as}
\hspace{0.7cm} u\rightarrow \infty.
\eeq
Defining
\beq
W(u)=W_r(u)+W_s(u),
\eeq
where $W_r(u)$ and $W_s(u)$ are respectively the regular and the singular
part of $W(u)$, we have
\beq
W_r(u)=\frac{1}{3}\left( 2 V'(u)+V'(-u)\right).
\eeq
Furthermore, by analyticity considerations~\cite{Eynard:1992cn} 
(see also~\cite{Engquist:2003rn})
one can 
show that $W_s(u)$ has to fulfill the following cubic equation
\beq\label{cubic}
(W_s(u))^3-R_1(u)W_s(u)-R_2(u)=0,
\eeq
where
\bea
R_1(u)&=&4\pi^2 \left(m+\frac{\epsilon}{4\pi c}\right)^2+\frac{1}{3u^2}, 
\nonumber \\
R_2(u)&=& \frac{2}{27u^3}+\left(\frac{\alpha}{2}-\frac{1}{3}\right)
8\pi^2\left(m+\frac{\epsilon}{4\pi c}\right)^2 \frac{1}{u}.
\nonumber
\eea
Solving the equation~\rf{cubic} perturbatively for large $u$ we get a
relation between $\epsilon$ and $c$. It reads
\beq
\frac{1}{4\pi c}=\frac{m\epsilon}{\alpha(1-\frac{3}{4}\alpha)}.
\eeq
Next, solving the equation~\rf{cubic} perturbatively for small $u$ we
get an expression for $\gamma$ (cf.\ eqn.~\rf{gammacont})
\beq
\gamma=\frac{\lambda \alpha}{2L}\left(m+\frac{\epsilon}{4\pi c}\right)^2
\approx \frac{\lambda \alpha m^2}{2L} 
\left(1+\frac{2\epsilon^2}{\alpha(1-\frac{3}{4}\alpha)}\right).
\eeq
We can express $\alpha$ as
\beq
\alpha=1-\frac{J_1}{L}\equiv 1-j_1,
\eeq
which leads to the following expression for $\gamma$
\beq\label{gammapert}
\gamma=\frac{\lambda m^2}{2L} \left(1-j_1+8\epsilon^2
\frac{1}{1+3j_1}+{\cal O}(\epsilon^4) \right). 
\eeq
Using the formalism of~\cite{Arutyunov:2003uj} one can derive in parametric
form an expression for the semi-classical energy of a three-spin circular
string of elliptic type with winding number $m$. Using the same notation
for the angular momenta as above the result reads~\cite{Arutyunov:2004}
\beq
E=L+\frac{\lambda m^2}{2L}
\left[ \frac{4}{\pi^2} \frac{\mbox{K(t)}}{\mbox{E(t)}}
\left((\mbox{E(t)})^2+j_1(t-1)(\mbox{K(t)})^2\right)\right],
\label{energy}
\eeq
where $t$ is determined as a function of $\epsilon$ and $j_1$ from the
following equation
\beq
\epsilon =\frac{1}{t}-\frac{1}{2}-\frac{\mbox{E(t)}}{t\,\mbox{K(t)}}
+j_1\left[\frac{1}{t}-\frac{1}{2}-\frac{\mbox{K(t)}}{t\,\mbox{E(t)}}
+\frac{\mbox{K(t)}}{\mbox{E(t)}}
\right],
\label{teps}
\eeq
with K(t) and E(t) being the elliptic integrals of the first and
the second kind, respectively.
Solving eqn.~\rf{teps} for $t$ in terms of $j_1$ to leading order
in $\epsilon$ and inserting the solution in eqn.~\rf{energy} one finds that
to the given order in $\epsilon$ the $\lambda$-dependent part of $E$ 
precisely agrees with the expression for $\gamma$ in eqn.~\rf{gammapert},
i.e.\
\beq
E=L+\frac{\lambda m^2}{2L} \left(1-j_1+8\epsilon^2
\frac{1}{1+3j_1}+{\cal O}(\epsilon^4) \right). 
\eeq
Thus, we propose that the dual of the operator considered here is the 
three-spin circular elliptic string of~\cite{Arutyunov:2003uj}.
It would of course be interesting to reproduce the equations~\rf{energy}
and~\rf{teps} from an exact solution of the integral equation~\rf{intlargec}.

\section{Conclusion}
 We have studied a class of single trace scalar, holomorphic
gauge theory operators with general $R$-charge assignment
$(J_1,J_2,J_3)=((1-\alpha)L,(\alpha-\beta)L,\beta L)$ in the limit
$L\rightarrow \infty$ with $\alpha\in [0,\frac{1}{2}]$ and 
$\beta\in [0,\frac{\alpha}{2}]$.  
Analyzing the relevant Bethe equations we have exposed the
analyticity
structure of the problem of determining the one-loop anomalous
dimension
of these operators. In particular, we have 
located a line of critical points in the parameter space, 
$\beta=\beta_c(\alpha)$, which explains why the nature of the dual
string, as observed, does not need to be the same for $\beta\rightarrow 0$ and
$\beta\rightarrow\frac{\alpha}{2}$.
Furthermore, we have proposed that for $\beta>\beta_c(\alpha)$ the
gauge theory operators studied are the duals of the circular elliptic 
three-spin string of~\cite{Arutyunov:2003uj} and supported this
by a perturbative calculation. It would of course be interesting
to identify the dual string state also for $\beta<\beta_c(\alpha)$.
The only candidate available at the moment seems to be the hyper-elliptic
three-spin state of~\cite{Arutyunov:2003uj} which generalizes the
two-spin folded string of~\cite{Frolov:2003tu}. As we have seen there
exists a mechanism encoded in the Bethe equations which effectively
leads to the appearence of extra cuts but it seems that the Bethe root
configurations studied here are still not general enough to lead to
a true hyper-elliptic structure. In the integrable Neumann model the
hyper-elliptic structure is reflected by the appearence of two
integer winding number like parameters. The corresponding 
(but not identical) degrees
of freedom of the folded string are  the number of
foldings and the number of so-called bend points. The folded three-spin
rigid string of~\cite{Arutyunov:2003uj} needs to have at least one
bend-point.
 In the case of the two-spin folded string it is known that
the parameter $m$ in eqn.~\rf{rootone} counts the number of
foldings~\cite{Beisert:2003ea,Frolov:2003xy} but it is
not obvious how bend points would manifest themselves on the
gauge theory side. A detailed
understanding of the nature of the operators studied for 
$\beta<\beta_c(\alpha)$ and their relation to semi-classical string
states requires an exact solution of the integral equation~\rf{intsmallc}
and we hope to
report on this in the future~\cite{inprogress}.
An exact expression for the resolvent associated
with the density of Bethe roots $\{u_{1,j}\}_{i=1}^{n_1}$
would not only give us access to the
one-loop
anomalous dimension of our gauge theory operators but also to the 
infinite set of conserved higher charges~\cite{Arutyunov:2003rg}. In
this connection it should be mentioned that one might envisage
a more direct way of comparing gauge theory and string theory results,
namely by directly deriving the relevant string sigma model from the
spin chain. So far this has only been accomplished for the simple
case of the $SU(2)$ sub-sector of the $SO(6)$ integrable spin 
chain~\cite{Kruczenski:2003gt}. Another interesting line of
investigation
which has also only been pursued in a
sub-sector not including the operators considered here
is the derivation of the dilatation operator to higher loop
orders~\cite{Beisert:2003tq,Beisert:2003jb,Beisert:2003ys}
%, which has been done for $SU(2)$~
%\cite{Beisert:2003tq,Beisert:2003jb}
%and $SU(2|3)$~\cite{Beisert:2003ys}, 
and the formulation of the corresponding 
Bethe ansatz~\cite{Serban:2004jf}.
%which has been done for $SU(2)$~\cite{Serban:2004jf}.

\vspace*{1.0cm}
\noindent
{\bf Acknowledgements}

We thank Gleb Arutyunov, Stefano Kovacs, Jorge Russo,
Matthias Staudacher and Arkady Tseytlin 
for useful discussions. We are particularly grateful to
Gleb Arutyunov for sharing
with us the unpublished notes~\cite{Arutyunov:2004}.
We furthermore acknowledge the support of the 
EU network on ``Discrete Random Geometry'', grant HPRN-CT-1999-00161.

\end{document}